# The road to accuracy: machine-learning-accelerated silicon ab initio simulations

M. Sluydts[1, 2, 3]*, M. Larmuseau[1, 2, 4], J. Lauwaert[5], S. Cottenier[1, 2]

[1]*Center for Molecular Modeling, Ghent University, Belgium*

[2]*Department of Electrical Energy, Metal, Mechanical Constructions & Systems, Ghent University, Belgium*

[3] *ePotentia Data Science and IT services, Antwerp, Belgium/Largo, FL, United States*

[4]*Department of Information Technology, Ghent University - imec, Belgium*

[5]*Department of Electronics and Information Systems, Ghent University, Belgium*

e-mail: michael.sluydts@ugent.be, stefaan.cottenier@ugent.be

**Abstract**

Ab initio simulations are capable of providing detailed information of material behavior at the nanoscale. Simulating experimentally relevant situations is, however, often computationally intense. Using hybrid approaches between ab initio methods such as density functional theory (DFT) and machine learning, new models can be constructed which retain quantum accuracy while being computationally faster by several orders of magnitude. Two examples are discussed in this paper. The first is the computational search for low energy substitutional defect complexes in silicon. The second is the construction of deep learning potentials for ab initio-level molecular dynamics simulations. The latter is applied to reproduce the 0 K equation of state and high temperature thermal expansion of Si using the same model.

**Introduction**

Many years of Si research have made it possible to create Si-based materials with a high degree of control over the material purity, included defects and resulting materials properties. The result is a revolution of intricate electronic devices, whose dimensions are now reaching down to the nanoscale. Operating at this scale requires updating our understanding of semiconductor physics to include quantum effects and may rely on detail down to the single atom level. At the same time, the high demands put on Si materials force us to continue improving our understanding of Si and its impurities even for macroscopic wafer production. Especially for defects it is, however, not guaranteed that all desired properties are readily accessible from experiment.

An ideal way to obtain high accuracy information about phenomena on the nanoscale is to construct the desired atomic configuration and explicitly simulate them using quantum mechanics. Ab initio methods have proven their worth in recent years[1], but suffer from one drawback: the more accurate the simulation, the more intense it becomes computationally. The result is a permanent trade-off between the number of combinations studied, the length- and time-scale of each individual simulation and the accuracy of the result with respect to experiment. To make these methods routinely useful for experiment this trade-off should be eliminated.

One possible way to avoid this deadlock is to rely on a new set of techniques from the field of machine learning. These rely on the fact that ab initio methods are highly generic. A good ab initio method can model equally well a disordered organic liquid as an inorganic ordered crystal. While this is magnificent from a physical perspective, it is one of the reasons why these methods are so resource-intensive. When interested only in the physical and chemical information related to an ordered Si crystal and its defects, the information included within the full laws of quantum mechanics is excessive and redundant. Machine learning models are capable of representing highly complex relationships between arbitrary inputs and outputs. The goal is to perform a sufficiently diverse range of Si ab initio simulations and then apply machine learning to extract only relevant information. The result is a model which retains ab initio accuracy, but at a computational cost reduced by several orders of magnitude.

In this paper we will illustrate recent advances in combining ab initio and machine learning methods through two examples. First, we will show how high-throughput screening, which tackles the traversal of large combinatorial spaces, can be accelerated. Secondly, we will create a deep learning potential for silicon. Reminiscent of a classical potential, it can be used to perform molecular dynamics simulations. In contrast to its classical predecessors however, a deep learning potential retains quantum mechanical information, allowing it to arbitrarily break bonds and undergo chemical reactions between species previously seen during the training procedure.

**High-Throughput screening**

The goal of high-throughput screening is to study trends in a large combinatorial space. The result is a large number of simulations, of which each individual simulation is necessarily limited in complexity. An example from is a database we have previously calculated containing properties of point defects and vacancy complexes in Si and Ge, a subset of which is shown in Figure 1.[2]

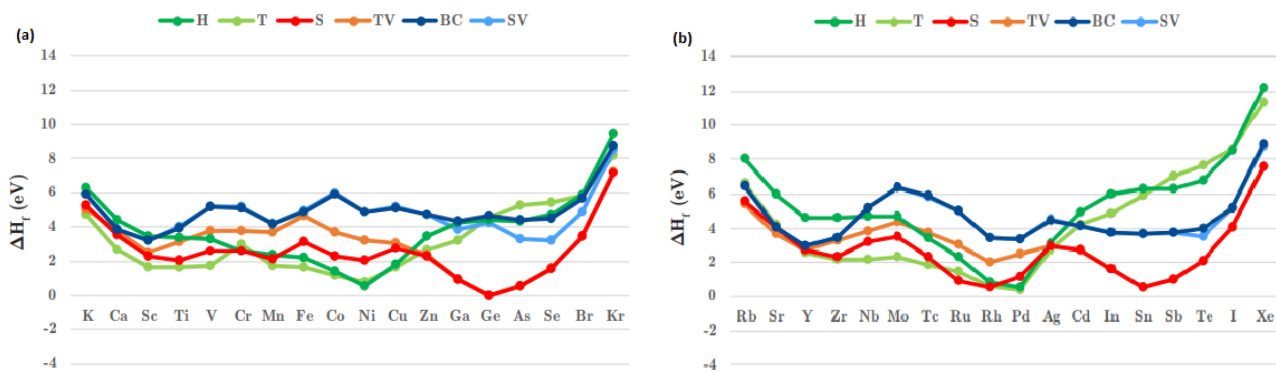

**Figure 1: Formation enthalpy of point defects and vacancy complexes in silicon for (a) period 4 and (b) period 5 impurity species in various configurations, calculated with hybrid density functional theory.**[2]

The subset includes the formation enthalpies of six point defects, for 36 elements of period 4 and 5. The result is 216 formation enthalpies, each of which required multiple calculations to obtain, even when ignoring temperature and charge effects. When confronted with this data, multiple interesting features can be identified. The first is that, while not identical, the curves are relatively similar between the two periods. This indicates an overlap of information, which a machine learning may be able to learn.

Secondly, all the curves are relatively smooth, suggesting that any point, when removed from the dataset, may be interpolated with reasonable accuracy from its neighboring points in the periodic table. When these two observations are combined, it strongly suggests it may not be necessary to calculate the entire dataset to obtain the information to create this curve. Moreso, often we may not be interested in the curve itself, but in a specific point. A typical example would be the defect with the lowest formation energy. In this situation, it is entirely unnecessary to obtain the full dataset with the same precision. Rather the key is to identify the interesting subspace of the search set using simulations of lower numerical precision or experimental accuracy and then performing the more intense calculations only in the region of interest.

**Intelligent Screening**

The goal is now to move from a brute-force high-throughput screening approach, where all combinations are simulated, towards an intelligent screening approach where only combinations containing essential information are studied. The way we will do this is using surrogate modeling[3], where a simple machine learning model will gradually replace the much heavier density functional theory-based model during the simulations. The example we will study is that of two adjacent substitutional defects in a 64-atom cell of pure silicon. The species will be selected from 72 elements from H to Rn, excluding the lanthanides other than La and Lu. The final number of unique combinations is the number of combinations of 2 elements out of 72 = 2556 as exchanging the two defects leads to the same crystal due to the symmetry of the silicon crystal. The resulting dataset is much larger than that shown in Figure 1, making a brute-force approach unfeasible.

For each defect complex, the formation energy with respect to all known and calculated stable crystals, the hull energy, is calculated using density functional theory.[4] This is a generalization of the formation energy in Figure 1, which is calculated only with respect to the pure phases. Each calculation is guided through an automated workflow using our automation software Queue Manager, which relaxes the geometry and volume of the cell and gradually increases the numerical precision.[2] The target we chose was to find the complexes most likely to occur, i.e. having a low hull energy. To this end we used a hull energy cutoff of 0.05 eV/atom, corresponding to a formation energy of the complex below 3.2 eV in the 64-atom cell. The next step is to perform an initial homogeneous sampling of the compositional space to serve as unbiased input data for an initial model. The size of this initial sampling is approximately 7% of the total set.

Then a Gaussian process model[5] is trained to predict the hull energy on this data. There are multiple reasons for choosing a Gaussian process model. It is capable of regression between numerical inputs and outputs and is fairly fast, which is essential if we want to gradually update it during the screening procedure without causing undesirable computational overhead. Most importantly, it provides for each prediction also a standard deviation as a means of uncertainty. The latter can be used to decide where the model needs additional sampling and how precise the predictions actually are. As input only atomic information is used, which can be obtained from the periodic table. Geometric information can be omitted as the geometry is virtually identical in all cases. Geometric stress and relaxation will lead to a modification of the hull energy, but this is considered a trend that the model must learn from the hull energy data. Once the first model has been trained, it predicts which defect complexes are most likely to satisfy the hull energy criteria and updates the priority of the calculations in Queue Manager appropriately. Each time a calculation finishes, the model is updated with new information. This decreases the uncertainty on the predictions, which then biases the search more towards the desired complexes. The predictive model has been trained on a fully calculated subset of approximately 10% of the full dataset and another 5% partially calculated data. Results are shown in Figure 2.

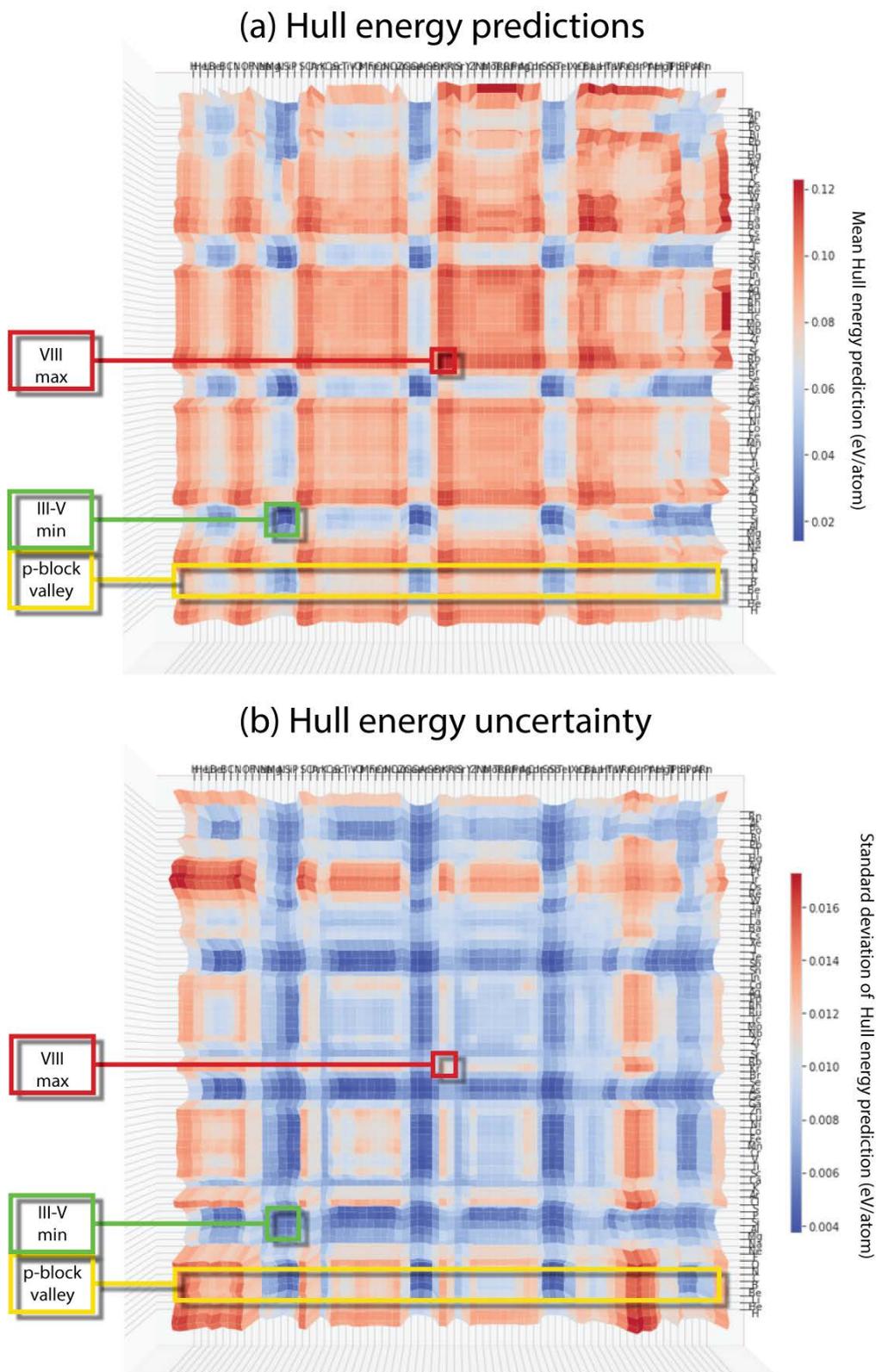

Figure 2: Gaussian process results for the hull energy of substitutional complexes in Si. (a) shows mean predictions, (b) shows the standard deviations on these predictions

Figure 2(a) shows the mean hull energy prediction made by the model, while Figure 2 (b) shows the standard deviation on each point of the hull energy surface. Important features have been noted on both figures. The surface spanned by the substitutional complex hull energies consists of a series of valleys with rectangular plateaus in between. The valleys represent the p-block, where it was previously seen in Figure 1 that the preferred defect position is substitutional, forming a parabolic formation energy curve around the group IV element of that period. The model has discovered the global minimum, which unsurprisingly is Si-Si itself, forming the absolute zero. All other IV-IV combinations form secondary local minima with a clear preference for combinations in the III-V region. The maxima are, also unsurprisingly, combinations of noble gases. The current global maximum is a combination of Rn and Kr, which is close to the value of a Kr-Kr pair noted on the figure. Combinations with noble gases form a high energy band on the plateau, together with the s block elements. The latter are both incapable of compensating the four broken Si-Si bonds in the substitutional position and quickly become large, inducing lattice strain.

If we investigate the uncertainty on the predictions in Figure 2(b), we see that the minimal energy regions are also the minimal uncertainty regions, as the model prioritizes calculations in these regions. Uncertainty in other regions is however acceptable provided it is sufficiently unlikely to change which complexes are most stable. The worst predictions occur for the second-period elements when combined with either themselves or the sixth-period transition metals. Second row elements are too small to appropriately occupy a substitutional position, let alone occupy a divacancy. The uncertainty may however have a secondary cause, namely that stricter numerical settings are required for the calculations, as well as a longer relaxation time to find a good geometry. The result is that collecting enough data takes longer and, while at the same time it is less likely to lead to a stable complex. When such higher uncertainty regions are encountered it may be worth additional sampling to ensure no energetic minima are missed. The noble gases are the next least certain, because they are least likely to become low in energy even after reducing the uncertainty.

The final result is a model which is capable of providing us with the requested information, the most stable complexes, while still giving us a generic idea of the overall behavior of the entire dataset while only having to calculate 10-15% of the full dataset.

**Deep learning potentials**

For the second application, we will focus on the mechanical properties of bulk silicon. These can be obtained by reproducing the full potential energy surface as provided by density functional theory calculations. Not only should this implicitly contain the 0 K equation of state but using molecular dynamics techniques it should also reproduce high temperature data which is notoriously resource-intensive using only density functional theory. Whereas the previous model only had to map the atomic identity of the defects to the hull energy, now the geometric position of each atom in the unit cell is important and must be considered. Such input is inherently more complex and thus requires both a more complex model architecture as well as significantly more training data. The chosen model comes from the class of deep learning models which, by propagating information through many layers, are capable of selecting and constructing important features from raw input data.

The specific architecture chosen is called a SchNet[6] architecture, which represents the silicon unit cell as an interconnected graph, of which each node represents an atom and each edge an interaction between two atoms. The model performs a graph convolution, i.e. it iterates over each atom in the crystal's cell. For each atom it constructs an environment within a certain cutoff radius, typically 5 – 8 Å. This environment is then described using a radial basis set, creating a rotationally invariant representation. The latter is preferred, as the rotation of a crystal or molecule should not influence the chemical and physical laws which it must adhere too. The SchNet model then uses several so-called interaction blocks to further transform and condense the representation of each atom and its environment in a way that is capable of capturing the energetic contribution they bring to the crystal. Finally, these energies are combined to make a prediction of the energy, which is then differentiated to obtain the forces. This differentiation step requires a smooth predicted potential energy surface, which is guaranteed due to the way the SchNet model is constructed.

The dataset used to construct the model was created from ab initio molecular dynamics using the VASP software using a 64 atom bulk Si cell.[7] The ab initio equation of state was created at 0 K using seven volume points (a = 5.35 – 5.53 Å). The seven volume points were then used as the starting point of three molecular dynamics (MD) simulations at 300 K, 1,000 K and 2,000 K respectively at constant volume and temperature (NVT).[8] Each molecular dynamics simulation ran for 500 steps with a 3 fs timestep. The result is a dataset of 10,500 datapoints. The input data used from each MD step is the 0 K internal energy, the forces on each atom and the cell parameters. Running high temperature simulations ensures that geometrical configurations which are high in energy are also sampled, while the low temperature data biases the dataset more towards the lower energy configurations. The model is trained using a custom implementation of the SchNet architecture.[9] After training, the resulting model is then used to perform static evaluations of geometries, as well as molecular dynamics simulations using the YAFF software package.[10] Results are shown in Figure 3.

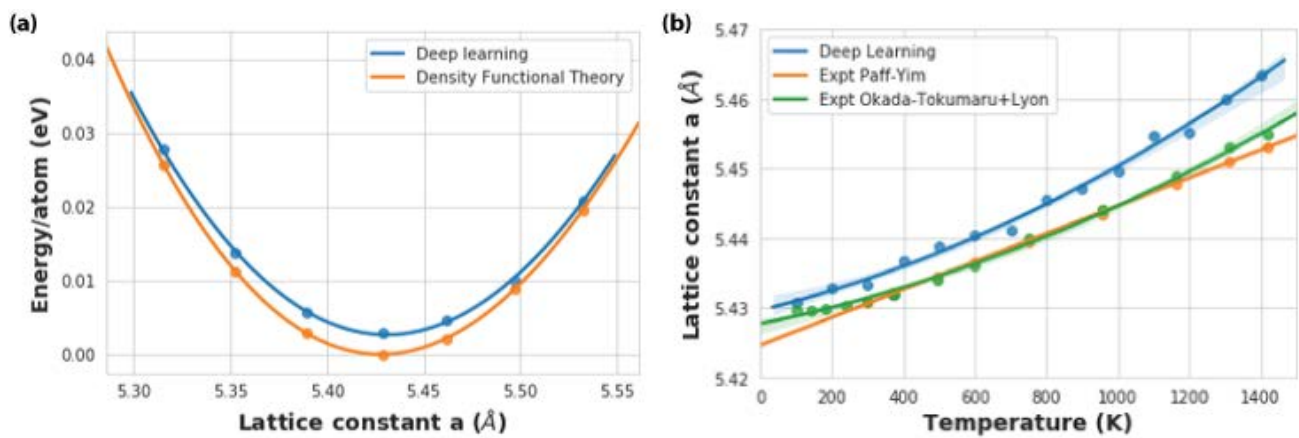

**Figure 3**: Evaluation of SchNet model for a (a) 0 K equation of state, compared with DFT (b) thermal expansion, compared with experiment.[11], [12]

Figure 3(a) shows the 0 K equation of state for bulk silicon calculated with both the SchNet deep learning model and density functional theory. The energy scale corresponds to absolute energy per atom as returned by the DFT calculation, referenced to the minimum of the DFT equation of state. The maximum energetic deviation of the model is slightly less than 3 meV/atom which is very small and similar to the difference of energies calculated between different implementations of the same DFT method.[13] The maximal deviation occurs around the absolute minimum which may be an artefact from the fact that our dataset only includes geometries from higher temperature molecular dynamics runs. As a result, the perfect bulk crystal structure is rarely encountered within the DFT molecular dynamics simulations. The consequence is that the energy will never drop as low as the absolute minimum. If we look at the obtained equilibrium lattice constant, we see that, despite the deviation in energy, the position of the minimum is virtually identical between DFT and SchNet, 5.428 and 5.43 Å respectively, which is also the same as the thermally-corrected experimental result of 5.43 Å.[14] The errors do affect the curvature of the equation of state and, consequently, the bulk modulus. The DFT value of 99.69 GPa is a close match to the thermally corrected value of 98.8 GPa, while SchNet gives a bulk modulus of 94.52 GPa.[14] The difference is still acceptable, but could likely be improved by sampling more near-equilibrium structures.

In Figure 3(b) the same equilibrium lattice constant is shown in function of temperature. These values have been obtained by performing constant pressure and temperature (NPT) MD simulations at each temperature and averaging the lattice parameter over the last 10,000 steps. Since this is an average over non-equilibrium structures, the resulting lattice constants are more noisy than the energy evaluations of the perfect crystal in the previous paragraph. The results are compared to experimental data by Yam/Piff and Okada/Tokumaru.[11], [12] Only the latter reports also data below 300 K, where thermal expansion is negligible and even temporarily negative. Between 300 and 1000 K the difference between SchNet and experiment is nearly constant with a value of 0.005 Å, suggesting the thermal expansion itself is described okay, but the volume is slightly overestimated which may be due to the reference ab initio data. This consistent behavior suggests the model is quite reliable in this temperature range. Above 1000 K, deviation from experiment however increases. The reason for this is the approach to the melting temperature of Si: 1687 K. While the SchNet has been trained on molecular dynamics simulations before and after the phase transition (1000 K and 2000 K), both started from an ordered crystal. Due to the length of the molecular dynamics simulations, the 2000 K crystal will not have had time to melt and will be more representative of a superheated perfect crystal.[15], [16] The latter likely continues to expand, before eventually losing its crystal structure and becoming a denser liquid. As the model has not seen this melting procedure it is likely interpolating an artificial thermal expansion in the 1000-2000 K temperature range. This can be corrected by adding additional data from molecular dynamics simulations before the melting point in the crystalline phase and appropriate liquid data above the expected phase transition.

**Conclusions and outlook**

The performance of mixed approaches using both density functional theory and machine learning techniques was investigated by means of two applications. In the first application, high-throughput density functional theory screening was used to search for the most stable adjacent substitutional defect complexes. A simple Gaussian process model was trained automatically during the screening procedure using atomic inputs and was used to direct the calculations towards the most likely candidates. As a result, a prediction could be made for the entire search space with only 10% of the materials fully calculated. The uncertainty on these predictions could also be quantified thanks to the Gaussian process model and suggested good performance in the desired regions. The second application concerned a more complex model capable of returning energy and forces for an arbitrary geometry. It was trained on density functional theory molecular dynamics simulations and then evaluated with respect to density functional theory and experiment. The model was capable of reproducing the 0 K equation of state obtained from density functional theory with a small error comparable to numerical noise observed between different implementations of the same method. The same model was used to perform constant pressure molecular dynamics simulations at a wide range of temperatures. The resulting lattice constants were then compared directly to experiment. Between 300 and 1000 K the lattice parameter was slightly overestimated, with a constant value. This suggests an appropriate description of the thermal expansion, with a constant deviation which may be inherent to the original density functional theory data. At higher temperatures the deviation increased, likely due to a lack of additional data before Si's melting point at 1687 K. A versatile model at high temperatures may also need to be exposed to data in the liquid phase. Nonetheless we see that even with little data a reasonably accurate deep learning potential can be trained. While the speedup in the initial application was only a factor ten, the deep learning potential can be evaluated around 5,000 times faster on a single GPU. Given sufficient data, these potentials should become similarly accurate as density functional theory and may eventually help us to routinely perform simulations at industrially relevant length and time scales on a single laptop.